# Stability of wurtzite semi-polar surfaces: algorithms and practices

Yiou Zhang, Jingzhao Zhang and Junyi Zhu[1]

Department of physics, the Chinese University of Hong Kong, Shatin, N.T., Hong Kong, China

**Abstract**

A complete knowledge of absolute surface energies with any arbitrary crystal orientation is important for the improvements of semiconductor devices because it determines the equilibrium and nonequilibrium crystal shapes of thin films and nanostructures. It is also crucial in the control of thin film crystal growth and surface effect studies in broad research fields. However, obtaining accurate absolute formation energies is still a huge challenge for the semi-polar surfaces of compound semiconductors. It mainly results from the asymmetry nature of crystal structures and the complicated step morphologies and related reconstructions of these surface configurations. Here we propose a general approach to calculate the absolute formation energies of wurtzite semi-polar surfaces by first-principles calculations, taking GaN as an example. We mainly focused on two commonly seen sets of semi-polar surfaces: a-family ($11\bar{2}X$) and m-family ($10\bar{1}X$). For all the semi-polar surfaces that we have calculated in this paper, the self-consistent accuracy is within 1.5 meV/Å$^2$. Our work fills the last technical gap to fully investigate and understand the shape and morphology of compound semiconductors.

In the past decades, despite the rapid development in wurtzite (WZ) based semiconductors that led to broad industrial applications, [1-6] high quality crystal growth and nanocrystal morphology controlling are challenging [7-10]. And the display technology based on quantum dots light-emitting diodes (LEDs) is towards the dawn of commercialization. [11,12] The crystal growth and morphology of group-III nitrides drew special interest because of the great success in the InGaN based LEDs [13]. Currently, the GaN-based optoelectronic devices are mainly limited to blue emitters on polar GaN grown on a c-plane (0001) sapphire [1-4,14]. And, it is difficult, if not impossible, to fabricate high-efficiency green- and yellow-light LEDs based on high quality InGaN with a high indium concentration,[10,14-17] due to the miscibility gap resulting from the large size mismatch between gallium and indium atoms and the piezoelectric effect on the polar surfaces. Growing GaN thin films along a semi-/non-polar direction, especially, semi-polar orientations, can be a promising approach to solve this fundamental obstacle, [18-25] because possible tensile sites on semi-polar surfaces can help the incorporation of large indium atoms [10,21]. In addition, semi-polar surfaces have relatively weak piezoelectric effects. [26] These lead to both enhanced indium incorporation [14,21,22] and reduced quantum confined Stark effects [26-28]. However, unlike the polar surfaces, fundamental understandings of semi-polar surfaces are largely missing because there lacks a working algorithm to estimate the absolute formation energy of semi-polar surfaces.

Absolute surface energy is one of the key quantities in surface studies.[9,29-32] A complete set of knowledge of absolute formation energies of GaN surfaces with all possible orientations is necessary for the related thermodynamic stability analyses, for example, determining the equilibrium crystal shape (ECS) by Wulff's theorem [33]. Such thermodynamic property is one of the key factors in understanding and controlling the growth of GaN nanostructures, which are considered as the major candidates for realizing broadband and multi-color emission [34-41].

For the ease of reading, we give a brief introduction to the definition of semi-polar surfaces.

---

[1] jyzhu@phy.cuhk.edu.hk

Semi-polar surfaces were firstly defined by Baker et. al. [42] as those planes with a nonzero $h$ or $k$ or $i$ index and a nonzero $l$ index in the ($hkil$) Miller–Bravais indexing convention, extending diagonally across the hexagonal unit cell and form a non-orthogonal angle with the c-plane.

Recently, we developed accurate and efficient approaches (passivation scheme) in the calculations of absolute surface energy of WZ polar surfaces [9,31,32], namely (0001)/(000$\bar{1}$). A working algorithm for semi-polar surfaces is missing. Li et. al [30] attempted to understand GaN crystal shapes by using a WZ wedge scheme. However, their results are rather approximated and no absolute surface energy was obtained. The total numbers of the surface configurations are incomplete. Also, no passivation of the slab was performed and unphysical charge transfer should be expected. In another work, only average surface energies of two conjugated surfaces were obtained [43]. Therefore, such practices may not be accurate enough to assess important crystal growth phenomena or predict crystal shapes correctly.

To obtain accurate absolute formation energy of semi-polar surfaces is especially challenging due to three reasons: (1) because of the asymmetric nature of the top and bottom semi-polar surfaces in a computational slab, it's difficult to evaluate each semi-polar surface individually; (2) a popular approach in early polar surface studies is to transfer the surface into a wedge or a cluster that mimics it, however, it's almost impossible to construct computationally small enough clusters or wedges for all semi-polar surfaces; (3) due to the step nature of the semi-polar surfaces, it's difficult to directly passivate the bottom surfaces with pseudo hydrogen (pseudo-H) atoms with known pseudo chemical potentials (for details of pseudo chemical potentials (PCPs) of pseudo-H atoms, please read ref. [31] ), without introducing large steric effects and unphysical stress, which destroy the accuracy of passivation scheme [9,31], to the slab.

To solve the above problems, the general principle of the algorithm design should be changed. Unlike all the previous practices in polar surface calculations, since it's impossible to directly passivate the bottom surface without severe steric effects, **the only logical and possible alternative approach is to modify the bottom semi-polar surfaces of the slab and cut them into surfaces that can be conveniently estimated by passivating proper pseudo-H atoms.** However, such cutting and passivation may also result in steric effects at the corner of the cut facts due to the step nature of the cutting scheme, although on the facets, such effects can be largely reduced. Therefore, special treatment on the estimation of the steric effects near the corner of the steps should be taken. **These new design principles are significantly different from any early attempts in the estimations of surface energies and can be generally applicable in the stability studies of many difficult surface orientations.**

In this paper, taking GaN as an example, we propose a general approach to calculate the absolute surface energies of semi-polar surfaces of WZ materials. Our first-principles calculations were based on Density Functional Theory [44,45] as implemented in VASP code [46], with a plane wave basis set [47,48] and PBE Generalized Gradient Approximation (GGA) as the exchange-correlation functional [49]. The slabs were separated by a vacuum of at least 15Å. All the atoms in the slab were allowed to relax until forces converged to less than 0.005eV/Å. The energy cutoff of the plane-wave basis set was set to 500 eV. We have done careful convergence tests for energy cutoff, K point and slabs' thickness. We focused on two commonly seen sets of semi-polar surfaces: a-family ($11\bar{2}X$) and m-family ($10\bar{1}X$).

For the aforementioned reasons, we use slab models with a modification of bottom surfaces in all the semi-polar surface calculations. One of two surfaces of the slab are usually passivated to avoid unphysical charge transfers. A direct passivation of bottom surface with pseudo-H atoms is almost impossible to both satisfy the electron counting model (ECM) [50] and avoid severe steric effect. Therefore, we modified the bottom surface of the semi-polar slabs into a zigzag structure, consisting of only polar and non-polar surfaces, which can be easily passivated without the steric effect, as shown in **Fig. 1(a)** and **1(b)**. And the top surfaces are those we try to investigate and calculate. The left and right sides of the slabs in **Fig. 1(a)** and **1(b)** should still conserve the periodic boundary conditions. Therefore, the absolute surface energy of the top surface can be calculated from:

$$\sigma_{\text{top}} = \frac{1}{\alpha}\left[E_{\text{slab}} - n_{Ga}\mu_{Ga} - n_N\mu_N - \sum \hat{\mu}_{H_{Ga}} - \sum \hat{\mu}_{H_N}\right], (1)$$

where $\mu_{Ga}$ and $\mu_N$ are the chemical potentials of the gallium and nitrogen elements, $\hat{\mu}_{H_X}$ ($X = Ga, N$) is the PCPs of the corresponding pseudo-H atoms. The summation symbols are kept since the PCPs would depend on the local electronic environment of the pseudo-H atoms. Further, if we assume a thermodynamic equilibrium between the bulk material and surface, we can write

$$\mu_{Ga} + \mu_N = E_{GaN} = E_{Ga} + E_{N_2} + \Delta H_f(GaN), (2)$$

where $E_{Ga}$ and $E_{N_2}$ are the total energies per atom of solid Ga and $N_2$ gas, and $\Delta H_f(GaN)$ is the formation enthalpy of WZ GaN. This would add restriction to **Eq. (1)**, and we can rewrite it as:

$$\sigma_{\text{top}} = \frac{1}{\alpha}\left[E_{\text{slab}} - n_{Ga}\left(E_{Ga} + \Delta H_f(GaN)\right) - n_N E_{N_2} - (n_N - n_{Ga})\Delta\mu_N - \sum \hat{\mu}_{H_{Ga}} - \sum \hat{\mu}_{H_N}\right], (3)$$

where $\Delta\mu_N = \mu_N - E_{N_2}$ is the relative chemical potential ($\Delta H_f(GaN) \leq \Delta\mu_N \leq 0$). Based on our previous work, if the pseudo-H atoms have enough space to relax, we can make use of the PCPs obtained from the tetrahedral clusters [9,31,32], which are cut from a zinc-blende GaN crystal. For pseudo-H atoms away from the intercepts between non-polar and polar surfaces, the local electronic environment is similar to that on the face of the tetrahedral clusters ($\hat{\mu}_{H_X}^{face}$). Surface atoms on the convex intercepts are usually attached to two pseudo-H atoms, which are similar to atoms on the edge of the tetrahedral clusters ($\hat{\mu}_{H_X}^{edge}$). Special attention must be put on the concave intercepts, since the pseudo-H atoms may interact and become steric. Additionally, we have done a convergence test for sizes of the zigzag structures shown in Fig.1, by enlarging the size of the zigzag caves. For example, the ($10\bar{1}2$) surface, we tried to use two zigzag structures of different sizes as shown in Fig.1(c) and (d). The zigzag caves shown in Fig.1(c) and (d) are the two possible configurations we calculated (it's possible to construct even larger structures, however, they are beyond our computing capability). It turns out that the absolute surface energy difference is within 1.2 meV/Å$^2$, which is reasonable small enough because such a difference is within our self-consistent accuracy of the algorithm, as shall be shown later. Also, the self-consistent accuracy of absolute surface energies may mainly depend on the accuracy of PCPs, which suggest that configurations with fewer pseudo-H per unit area lead to better accuracies. Therefore, all of the results in this paper are obtained based on the calculations by using the zigzag structures with the similar size of the one shown in Fig.1(c). Detailed bottom surface structures for the calculations of each surface in this paper are summarized in supplementary material.

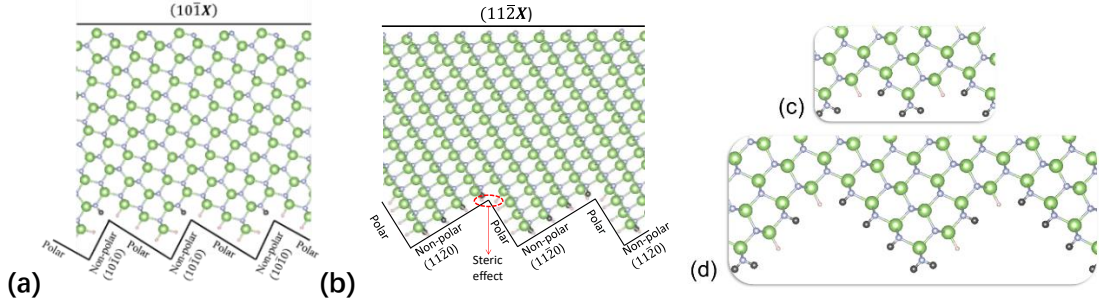

Fig. 1. (color online) Two representative bottom surface cuttings for semi-polar surfaces belonging to (a) m-family and (b) a-family. Large green balls represent Ga atoms, and small white balls represent N atoms, while small pink and black balls represent two types of pseudo-H $H_{Ga}$ and $H_N$. Dangling bonds of the atoms on the zigzag structure are passivated by corresponding pseudo-H atoms. For m-family, $(10\bar{1}3)$ surface is taken as an example. For a-family, $(11\bar{2}2)$ surface is taken as an example. (c) different sizes of the zigzag structures used for convergence testing.

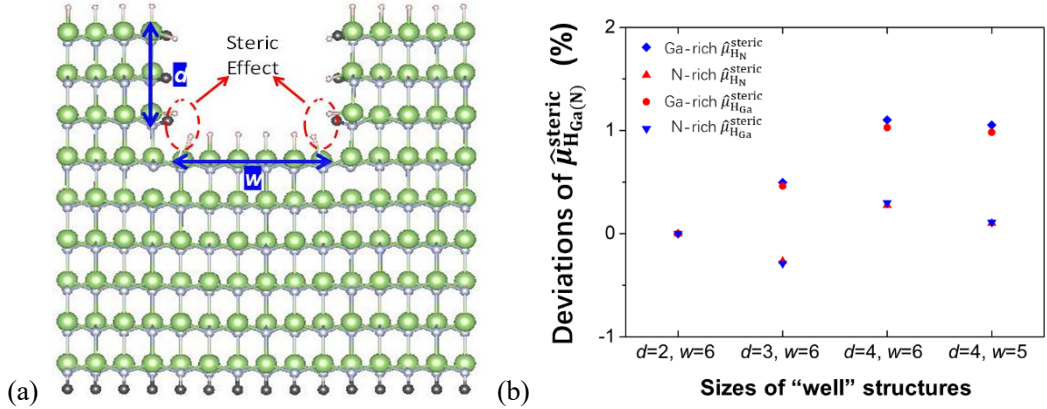

Fig.2 (color online) (a) Schematic diagram of "well" structures that mimic the steric effects on $(11\bar{2}X)$ surfaces. (b) Convergence test of $\hat{\mu}_{H_{Ga(N)}}^{steric}$ obtained from the "well" structure with different width (w) and depth (d), denoting the number of Ga/N atoms, and different choices yield similar values of the PCPs (difference within 1.1%, set the values with the size of d=2, w=6 as the reference zero).

TABLE 1. The calculated PCPs of steric pseudo-H with different sizes of "well" structures. Unit is eV.

|  | SIZES | | | |
| --- | --- | --- | --- | --- |
|  | d=2, w=6 | d=3, w=6 | d=4, w=6 | d=4, w=5 |
| Ga-rich $\hat{\mu}_{H_N}^{steric}$ | -3.176 | -3.192 | -3.211 | -3.210 |
| N-rich $\hat{\mu}_{H_N}^{steric}$ | -3.010 | -3.002 | -3.018 | -3.013 |
| Ga-rich $\hat{\mu}_{H_{Ga}}^{steric}$ | -3.410 | -3.426 | -3.445 | -3.444 |
| N-rich $\hat{\mu}_{H_{Ga}}^{steric}$ | -2.776 | -2.768 | -2.784 | -2.779 |

For the $(10\bar{1}X)$ surfaces, pseudo-H atoms on the concave intercepts are well separated, with the H-H interatomic distance larger than 2.5 Å. Therefore, PCPs obtained from the tetrahedral clusters are enough to determine the absolute surface energy:

$$\sigma_{(10\bar{1}X)} = \frac{1}{\alpha_{10\bar{1}X}} \Big[ E_{\text{slab}} - n_{Ga}\left(E_{Ga} + \Delta H_f(GaN)\right) - n_N E_{N_2} - (n_N - n_{Ga})\Delta\mu_N$$

$$- n_{H_{Ga}}^{face}\hat{\mu}_{H_{Ga}}^{face} - n_{H_{Ga}}^{edge}\hat{\mu}_{H_{Ga}}^{edge} - n_{H_N}^{face}\hat{\mu}_{H_N}^{face} - n_{H_N}^{edge}\hat{\mu}_{H_N}^{edge} \Big]. \quad (4)$$

The superscript (face or edge) indicates the approximated local electronic environment of the pseudo-H atoms. On the other hand, for $(11\bar{2}X)$ surfaces, pseudo-H atoms at the corner intercepted by non-polar $(11\bar{2}0)$ and polar $(0001)/(000\bar{1})$ surfaces demonstrate steric effects, as the distance between them is less than 2 Å. Our calculations indicate the pseudo-H are mutually repelled, as shown in Fig.1(b). The Ga-H bond angles are obviously distorted and induces large stress. To solve this stress problem, an intuitive approach is that the steric effects should be further simulated by constructing a similar computable structure. The new structure can be constructed by cutting a "well" on a conventional slab along c (or -c) direction, as shown in Fig. 2(a). The two corners of the "well" have the very similar H-H interatomic distances and Ga-H (or N-H) tilting angles as that of the steric corner of the bottom surface of the semi-polar slab, as shown in Fig.1(b). Therefore, this structure allows us to estimate the PCPs under steric effects:

$$\hat{\mu}_{H_{Ga}}^{steric} = \frac{1}{n_{H_{Ga}}^{steric}} \Big[ E_{\text{slab}} - n_{Ga}\left(E_{Ga} + \Delta H_f(GaN)\right) - n_N E_{N_2} - (n_N - n_{Ga})\Delta\mu_N$$

$$- n_{H_{Ga}}^{face}\hat{\mu}_{H_{Ga}}^{face} - n_{H_{Ga}}^{edge}\hat{\mu}_{H_{Ga}}^{edge} - n_{H_N}^{face}\hat{\mu}_{H_N}^{face} \Big]. \quad (5)$$

$\hat{\mu}_{H_N}^{steric}$ can be calculated in a similar way. According to our tests, the PCPs of the simulated steric hydrogen atoms are insensitive to the width and depth of the "well". Here we denoted the width of the "well" as *w*, and the depth of the "well" as *d*, Under different choices, difference between the calculated $\hat{\mu}_{H_{Ga(N)}}^{steric}$ values is within 1.1%, as shown in Fig.2(b), which indicates that such steric effects are localized and their propagation in the slab can be limited, unlike the early steric effects observed on the bottom surfaces of wedge structures [9]. Detailed data are summarized in Table.1. The localized feature of the strain energy is probably due to the concaved structure of the zigzag steps, which assist the stress relaxation. For wedge structures in early literature, such relaxations are relatively difficult due to the flat nature of the bottom (001) surfaces. These tests also suggest that although a direct modification of the bottom surface with zigzag structures may induce steric effects, such effects can be relatively localized and can be reasonably estimated by further simulations. Based on such PCPs, we can determine the absolute surface energies of $(11\bar{2}X)$ surfaces:

$$\sigma_{(11\bar{2}X)} = \frac{1}{\alpha_{11\bar{2}X}} \Big[ E_{slab} - n_{Ga}\left(E_{Ga} + \Delta H_f(GaN)\right) - n_N E_{N_2} - (n_N - n_{Ga})\Delta\mu_N$$

$$- n_{H_{Ga}}^{face}\hat{\mu}_{H_{Ga}}^{face} - n_{H_{Ga}}^{edge}\hat{\mu}_{H_{Ga}}^{edge} - n_{H_N}^{face}\hat{\mu}_{H_N}^{face} - n_{H_N}^{edge}\hat{\mu}_{H_N}^{edge} - n_{H_{Ga}}^{steric}\hat{\mu}_{H_{Ga}}^{steric}$$

$$- n_{H_N}^{steric}\hat{\mu}_{H_N}^{steric} \Big] . \quad (6)$$

While for group II-VI WZ compounds, such steric effects can be overcome through a real atom passivation, as shown in Fig.3. According to ECM, two anion dangling bonds lack one electron (Fig.3(a)) and two cation dangling bonds have one extra electron (Fig.3(b)). Therefore, group-I elements like lithium or group-VII elements like fluorine can be used to passivate these two anion or cation dangling bonds. The applicability of the similar passivation approach has been demonstrated in our previous paper on the bottom (001) surface of the wedge structure [9]. For group III-V compounds, such approaches are not applicable because of the fractional electron numbers. One possible way is to use fractionally charged elements [51] with proper atomic sizes, for instance, 1.5$e$ charged lithium used for passivating two nitrogen dangling bonds, or 6.5$e$ charged fluorine used for two gallium dangling bonds. Therefore, proper generated passivant pseudopotentials are necessary [52], while it is out of scope of this paper.

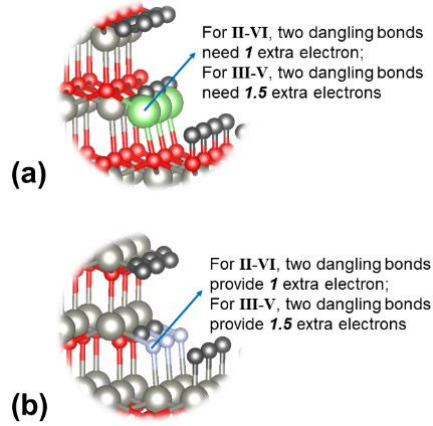

Fig.3. (color online) Illustrations of real atom passivation for group II-VI compounds for anion (a) and cation (b). Lithium, fluorine, group-II metal, group VI elements and pseudo-H atoms are denoted as green, light grey, grey, red and black atoms, respectively. This figure can also briefly demonstrate the alternative possible method solving steric effects for III-V compounds, where fractionally charged passivants are required.

Additionally, for the self-consistency (or accuracy) estimation, if both the top and bottom surfaces are cut into zigzag structures, and all the dangling bonds are passivated by pseudo-H atoms, we shall define a residue energy as:

$$E_{residue} = E_{slab} - n_{Ga}\left(E_{Ga} + \Delta H_f(GaN)\right) - n_N E_{N_2} - (n_N - n_{Ga})\Delta\mu_N$$

$$- n_{H_{Ga}}^{face}\hat{\mu}_{H_{Ga}}^{face} - n_{H_{Ga}}^{edge}\hat{\mu}_{H_{Ga}}^{edge} - n_{H_N}^{face}\hat{\mu}_{H_N}^{face} - n_{H_N}^{edge}\hat{\mu}_{H_N}^{edge} - n_{H_{Ga}}^{steric}\hat{\mu}_{H_{Ga}}^{steric}$$

$$- n_{H_N}^{steric}\hat{\mu}_{H_N}^{steric}. \quad (7)$$

If all the PCPs are exact, the residue energy should be zero. Therefore, we can use the residue energy to estimate the overall self-consistency of this computational approach. For all the semi-polar surfaces we have calculated in this paper, the residues are less than 1.5 meV/Å$^2$, which justifies the applicability of our method.

Briefly, the workflow of calculating the absolute surface energies of semi-polar surfaces are as following: (1) construct a conventional slab model with two conjugated semi-polar surfaces; (2) modify the bottom surface of the slab into corresponding zigzag structure, which should be passivated by proper pseudo-H subsequently; (3) obtain the PCPs of each pseudo-H used on the bottom surface according to the passivation scheme [9,31], and especially for a-family (11$\bar{2}$X) surfaces, estimate the PCPs of steric pseudo-H by "well" structures; (4) estimate the absolute formation energy of the targeted surfaces by subtracting the total energy of the slab with the corresponding chemical potentials and PCPs of all the slab atoms.

Unlike early algorithms, our new approach doesn't need to construct a wedge or cluster. Instead, we focused on the modifications of the bottom surfaces of the slabs and converted the unknown surfaces into surfaces that can be directly computed. Although we demonstrated this computational principle on the semi-polar surfaces of WZ GaN, it can also be applicable to other materials or surfaces. Our approach provides a powerful tool to investigate high index surfaces and the ECS of nanocrystals for other compound materials accurately.

By applying this method, in this paper, we have calculated (22$\bar{4}$1), (11$\bar{2}$1), (11$\bar{2}$2), (11$\bar{2}$3), and their conjugate surfaces, as well as (20$\bar{2}$1), (10$\bar{1}$1), (10$\bar{1}$2), (10$\bar{1}$3), and their conjugate surfaces. Strictly speaking, the most stable surface structures must be determined through exclusive search over all possible surface cuttings and reconstructions. Nevertheless, due to the low symmetry of semi-polar surfaces, it is difficult, if not impossible, to calculate all the possible cutting or reconstructions. To simplify the problem, we constructed the surface structures by cutting the bulk GaN crystal along the semi-polar directions. The cutting plane is moved up and down to cover all possible cuts. Along some semi-polar directions, the cutting surface can pass through two nonequivalent atomic sites, or the two sites can be quite close to the cutting surface. In such cases, we have considered surface structures with (1) one of the two sites included, (2) both sites included, and (3) neither site included. All of the surfaces configurations calculated in this paper are summarized in the supplementary material.

On (10$\bar{1}$X) surfaces, possible dimer formations have been considered explicitly. Also on (10$\bar{1}$1) surface, we have calculated the (4 × 2) Dimer-Vacancy reconstruction, which was reported as the most stable surface structure under N-rich condition [53]. Compared with the surface structures we have constructed, the (4 × 2) Dimer-Vacancy reconstruction is more stable by at most $10 \text{meV}/\text{Å}^2$ (under N-rich limit).

For comparison, we have also included the absolute surface energies of polar (0001)/(000$\bar{1}$) surfaces and non-polar (11$\bar{2}$0)/(10$\bar{1}$0) surfaces, with adatoms or adlayers surface reconstructions identified in previous literature [29]. Some Ga-rich surface reconstructions (e.g. Ga bilayer on (0001) surface) are not included. Such reconstructions are usually stable at Ga-rich limit. Nevertheless, since Ga-rich condition is not commonly seen in the real growth process, this would not largely affect our conclusions.

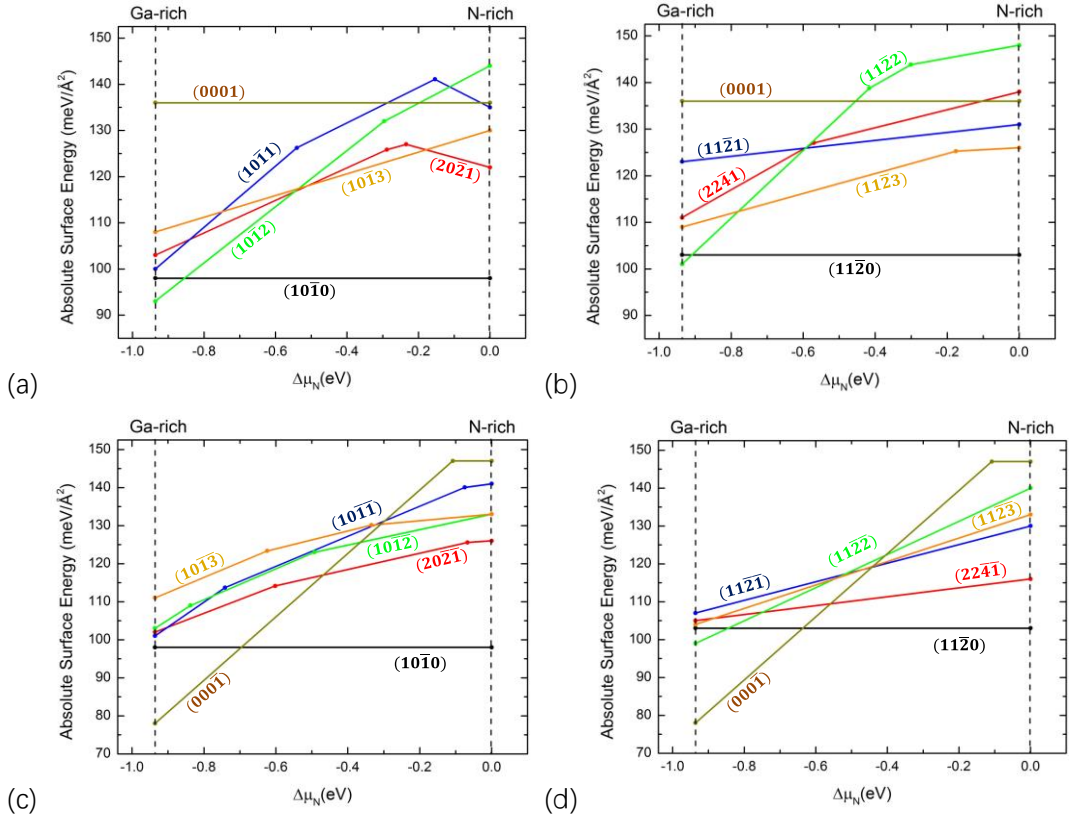

**Fig. 4.** (color online) Lowest surface energies of semi-polar surfaces under different chemical potential conditions. For clarity, we have divided the surfaces into four groups: (a) m-family in Ga-polar hemisphere, (b) a-family in Ga-polar hemisphere, (c) m-family in N-polar hemisphere, and (d) a-family in N-polar hemisphere. The polar and non-polar surfaces are also included for comparison.

Based on the obtained absolute surface energies of various surface structures, we have determined the lowest surface energies at different chemical potential conditions, as shown in Fig. 4. In general, the semi-polar surfaces have lower energies under Ga-rich condition. This is expected since dangling bonds of Ga atoms have lower energies and higher flexibility in rebinding. Additionally, a-family $(11\bar{2}X)$ surfaces generally have comparable or even lower surface energies than m-family $(10\bar{1}X)$ surfaces. Under extremely Ga-rich condition, $(10\bar{1}2)$ and $(11\bar{2}2)$ have the lowest energies. For the conjugated side, $(000\bar{1})$ dominates in a wide region near Ga-rich condition. And for $(10\bar{1}X)$ surfaces, their energies are very close except $(10\bar{1}3)$, while $(11\bar{2}2)$ has the lowest energy among a-family. Under extremely N-rich condition, nonpolar surfaces are the most stable ones. Except for that, $(20\bar{2}1)$, $(11\bar{2}3)$, $(20\bar{2}\bar{1})$ and $(22\bar{4}\bar{1})$ may dominate. Surprisingly, we can see from Fig. 4 that $(10\bar{1}1)$ and $(11\bar{2}2)$ surfaces, which are commonly seen facets of GaN crystal in experiments [35,43,54], do not have lower surface energies than the other semi-polar surfaces in a large chemical potential range (except for $(11\bar{2}2)$ surfaces under extremely Ga-rich condition). Also, $(10\bar{1}2)$ surface, which was also revealed in experiments [55,56], is the most stable one in the region near Ga-rich. While it can also be observed under N-rich condition in experiments. These experimental phenomena may mainly result from kinetic effects, like the Ga diffusion length [55] etc., and impurity passivation, as is also suggested by previous experiments [57]. All these affecting factors are out of the scope of this paper and may require further investigations.

In previous theoretical work [30] it has been concluded that ($10\bar{1}1$) has the lowest absolute surface energies under Ga-rich condition, and ($10\bar{1}2$) dominates under N-rich condition, after extending the chemical potential range by 2 eV on both sides. It also concluded that ($000\bar{1}$) dominates the conjugated side in the whole chemical potential region. Nevertheless, the passivation scheme in those previous works is subject to steric effects, and limited numbers of semi-polar orientations and few surface configurations have been taken into account. These could lead to rather large errors in the calculations. Meanwhile, if we considered the slopes of each surfaces in Fig.4 (a) and (b), when the chemical potential region is further extended to the left side similarly like previous works [30,58], mimicking experimental temperature and pressure conditions, ($10\bar{1}1$) and ($11\bar{2}2$) would be the most stable surfaces.

However, experimental studies have implied that the crystal shape of epitaxially grown GaN, especially nanostructures, depends more on non-equilibrium growth process.[43,54,59] Our calculations confirm that zero temperature absolute surface energies are not enough to explain crystal shape of GaN, and other factors like growth kinetics, surface impurity passivation etc., and temperature effects must also be considered. Generally, our results can explain the experimental facts [55,59,60], from thermodynamics point of view, that N-polar GaN nanowires are always terminated with ($000\bar{1}$) orientated flat tips, while Ga-polar ones show pyramid-like shapes with various orientations.

Since growth kinetics and impurity passivation also play important roles in determining crystal shape of GaN, it is essential to undertake the related researches on these important surfaces. Our method and results in this paper indeed provide a significant foundation for possible further investigations. This method here is demonstrated in WZ materials, while its general principles can be extended and applied to many other materials. Additionally, this approach also provides a chance to investigate the interfaces and growth mechanisms of heterostructures like ZnO/GaN, AlN/GaN along semi-polar orientations.

In summary, we have proposed a new method to calculate the absolute surface energies of semi-polar surfaces, and analyzed the growth behaviors of GaN crystal. Our new passivation scheme has mostly solved the steric effect and the accuracy in obtained absolute surface energies is within $1.5 meV/Å^2$. Based on this passivation scheme, we have calculated the surface energies of different surfaces under various surface structures. We have found that the lowest surface energies of the semi-polar surfaces are quite close. Hence, we theoretically confirmed that zero temperature thermodynamics is not enough to explain the crystal shape of GaN in experiments. Our results provide an efficient and accurate method to calculate the absolute semi-polar surface energies of other materials. Also, our works shall shed light on the determination of crystal shape thermodynamically, and provide a starting point for further kinetics researches.

**Acknowledgements**

Part of the computing resources was provided by the High-Performance Cluster Computing Centre, Hong Kong Baptist University. This work was supported by the start-up funding, HKRGC funding with the Project code of 2130490, and direct grant with the Project code of 4053233, 4053134 and 3132748 at CUHK.